\documentclass[pra,twocolumn,showpacs]{revtex4}
\usepackage{stmaryrd}
\usepackage{amssymb}
\usepackage{graphicx}
\usepackage{dsfont}
\def\>{\rangle}
\def\<{\langle}

\begin{document}

\title{Information-disturbance tradeoff in sending Direction information via antiparallel quantum spin }
\author{ShengLi Zhang$^{1}$, XuBo Zou$^{1}$, ChuanFeng Li$^{1}$,  ChenHui Jin$^{2}$
and GuangCan Guo$^{1}$}
\affiliation{1 Key Laboratory of Quantum Information, University of Science and
Technology of China (CAS), Hefei 230026, China.\\
2 Electronic Technology Institute; Information Engineering University;
Zhengzhou; Henan 450004; China}
\date{\today}

\begin{abstract}

For sending unknown direction information,  antiparallel spins contains more direction information than parallel spins( Gisin and Popescu, 1999, \textit{Phys. Rev. Lett.} \textbf{83}, 432).
In this paper, the optimal information-disturbance tradeoff
bound for antiparallel spins is derived. The quantum measurements which attain the optimal tradeoff bound are obtained.  
This result can be of practical relevance for posing some general limits on Eve's eavesdropping process. Finally, we also present a comparison between the bound for antiparallel spins and the bound for parallel spins.
%A strict bound for antiparallel spins, along with the optimal POVM measurement which attains the bound is obtained.

\end{abstract}

\maketitle

\section{Introduction}

With a preestablished reference frame between two remote users,
Alice and Bob, the information of an arbitrary spatial direction
$\vec{n}$ can be conveniently encoded into a series of classical
bits which allow them to exchange via quantum or classical channels.
However, there are many cases when such a reference frame is not
available and all what one can do is to send a natural object, such
as gyroscope pointing in a direction, to share and align their
reference frames\cite{RMP}. Quantum spins-1/2 systems, polarized in
direction $\vec{n}$, have been considered as a promising candidate
for processing such \textquotedblleft unspeakable\textquotedblright
information\cite{pereUnspeak}. Since the seminal work by Peres and
Wootters\cite{pere91}, a considerable effort has been made in the
literature to derive the optimal procedure in sending and receiving
frame
information\cite{Mass95,pere01,pere01carte,Nspin1,Nspin2,Nspin3,Gisin,Nspin4}.
For a single spin, the optimal encoding procedure is obvious and
straightforward\cite{Mass95}: Alice simply uses a spin pointing into
$\vec{n}$ to encode the direction, and the optimal measurement
method for Bob is a standard Stern-Gerlach measurement, along an
arbitrary direction $\vec{m}$. The measurement result $\pm m$
provides a fidelity of $2/3$ which is the maximal accuracy with
which Bob can achieve from the quantum measurement. In
Ref.\cite{Gisin}, Gisin and Popescu considered this transmission
problem using two spins and discovered a surprising effect which is
now often coined as \textquotedblleft Nolocal without
entanglement\textquotedblright. In more details, to transmit the
direction $\vec{n}$ using two spins, there may be two possible
strategies. The first one is to encode $\vec{n}$ in parallel spins
$|\vec{n}\rangle|\vec{n}\rangle$, while the second one is quite
similar to the first, except to polarize the second spin in the
opposite direction, $|\vec{n},-\vec{n}\rangle$ (antiparallel).
Although in both cases the two spins are unentangled, it is shown
that antiparallel quantum spin provides a definite improvement in
the precision and efficiency in the transmission of frame
information. Recently, the best strategy for efficient use of $N$
quantum spins to align the reference frame have also been addressed
\cite{Nspin4}.

%Moreover, in laboratory, optical
%experiments which achieves the optimal communication efficiency for
%2$\otimes$2 quantum systems was recently reported\cite{expFrame}.

All these studies of reference frame transmission is centered around
the improvement of the efficiency or the fidelity in our
communication. In the real world, particularly in the presence of
potential eavesdroppers (Eve), it is also of great importance to
keep the security of the shared frame. In Ref. \cite{Chir}, two
quantum-cryptographic protocols ------ \textit{BB84-type} protocol
and \textit{Ekert-type} protocol have been proposed for secretly
communicating a reference frame. However, up till now, a
quantitative derivation of the security level ( Bell inequalities)
above which the BB84-type (Ekert-type) protocol is no longer secure
has not been explicitly derived. In this paper, as a first step
towards such a goal, we take the antiparallel quantum spins which
provides the maximal transmission fidelity for two-qubit encoding as
an example and consider the corresponding security bound.

Unintuitively, however, as we will show here, the antiparallel spins
may not be the optimal protocol for transmitting the direction
information, at least from the aspect of security. Although
antiparallel spins provides Alice a convenient tool for improving
her fidelity for frame transmission, they improve the fidelity for
Eve, too. This allows Eve more freedom to eavesdrop. Thus, a more
theoretical and information theory-based analysis is required.

 Our result is
obtained by a careful derivation of the information disturbance
tradeoff problem. In fact, the laws of quantum mechanics imposes a
natural restriction on the information processing with the unknown
quantum state. There is not a quantum measurement on the quantum
system without introducing any disturbance. The more information one
gains, the more the quantum state has to be disturbed. There exists
a precise tradeoff between the information gain and state
disturbance. More importantly, the tradeoff which is inherited by
quantum mechanics is applicable to any measurement observer,
including Bob and eavesdropper, and imposes a general limit on the
information eavesdropping in quantum
communications\cite{fuchs96.pra,fuchs01.pra,banaszek01.prl,banaszek00.pra,
mista05.pra,discrimiVS,macca,M.F.Sacchi}.

In the following (Sec.\ref{model}), we will give a detailed description of the information-disturbance model for the security analysis of communication
protocol with antiparallel spins. In Sec.\ref{bound} we derive the optimal tradeoff bound by using group covariant and
vector analysis technique. Finally, an exemplary operation satisfying the tradeoff bound is constructed and
Sec.\ref{Conclu} follows the conclusions.

\begin{figure}
  \includegraphics[width=8.5cm]{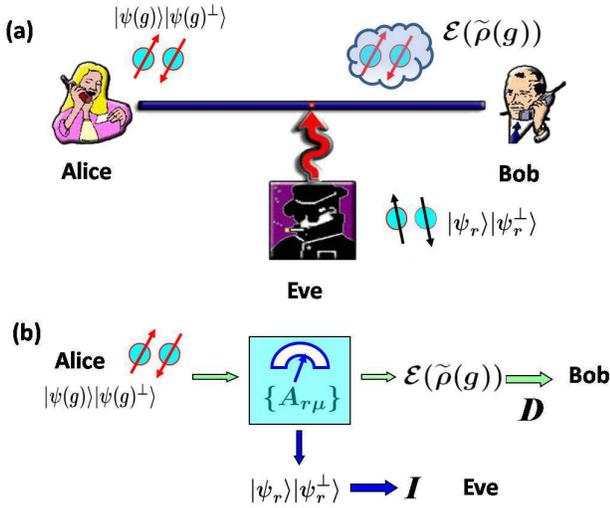}\\
  \caption{(a) Schematic illustration of Eve's wire tapping attack. (b) Mathematical model for the
  Information-Disturbance tradeoff.}\label{figscheme}
\end{figure}

\section{Information-Disturbance Model for security analysis}\label{model}

For convenience, let's present some basic notations that will be frequently utilized in the rest of the paper. Firstly, we will identify the desired
direction $\vec{n}$ which Alice wants to communicate with a group parameter $g\in\mathbb{G}=\mathrm{SU}(2)$. Actually, with respect to
a reference $\vec{n}_0=(0,0,1)$, there exists a rotation $g$ connecting the unit vector $\vec{n}$: $\vec{n}=g\vec{n}_0$. Thus, we can rewrite the
state
\begin{eqnarray}
|\vec{n}\rangle\rightarrow|\psi(g)\rangle\equiv U_g|0\rangle,\label{vecn}
\end{eqnarray}
with $U_g$ denoting the unitary group representation for $\mathbb{G}$ and $|0\rangle$ the corresponding quantum spin state for direction $\vec{n}_0: \vec{n}_0\cdot\vec{\sigma}|0\rangle=|0\rangle$
\cite{Nspin3}. Secondly, for what follows, we will also use the integral $\int (\cdot )d\vec{n} = \int (\cdot ) dg$ over the space of all possible pure state.
For ease, we consider the invariant and normalized Harr measure on $\mathrm{SU}(2)$, i.e., $\int dg=1$.

We shall now pull back our attention to the communication scenario.
Suppose the unknown direction information is stored and transmitted
with the antiparallel quantum state
$\widetilde{\rho}(g)=|\psi(g)\rangle\langle\psi(g)|\otimes|\psi(g)^{\perp}\rangle\langle\psi(g)^\perp
|$. After the quantum communication, the receiver, Bob, performs an
optimal measurement to retrieve it(Fig. \ref{figscheme}(a)).
However, in general, the quantum state received by Bob is not always
the original pure state, but a disturbed quantum mixed state
$\mathcal{E}(\widetilde\rho(g))$. There are many reasons for the
disturbance, some may come from the inevitable environmental noise
and decoherence process, whereas, others may be attributed to Eve's
active attack. For security, we conservatively conjecture that Eve
has a full control of the environmental noise and all the
decoherence is assumed to be due to eavesdropping. In Fig.
\ref{figscheme}(b), we give a mathematical model for Eve's
eavesdropping process. Without loss of generality, we here suppose
Eve performs a POVM measurement to carry out  his eavesdropping.
This is the most general measurement in quantum mechanics and can be
described with a collection of completely positive (CP) maps
$\{\mathcal{E}_r\}$, where $r$ denotes the possible measurement
results\cite{Davies,Kraus}. Moreover, choices of different map
$\mathcal{E}_r$ or of different number of distinct measurement
results will constitute different kinds of POVM measurement on
condition that the map $\sum_r\mathcal{E}_r$ satisfies the
trace-preserving condition. Namely, there may be infinite many
strategies for Eve to choose to maximize his eavesdropping
information. In the literature, many different definitions such as Shannon entropy\cite{macca}, discrimination probability
 \cite{discrimiVS} and the quantum fidelity\cite{banaszek01.prl,M.F.Sacchi}
have been used to quantify the amount of Eve's information. For
ease, throughout the paper, we will use the fidelity between the
state $|\psi_r\rangle$ (Eve guessed from his measurement result $r$)
and original state $|\psi(g)\rangle$ as a figure of merit for Eve's
information $\mathcal{I}$. After the measurement, the state will be
extensively disturbed.  The disturbed version, say
$\mathcal{E}(\widetilde{\rho}(g))$, will be transmitted to Bob for
subsequent processing. The amount of the disturbance $\mathcal{D}$,
characterized by the resemblance between the single qubit state and
original state $|\psi(g)\rangle$, i.e.,
$\mathrm{Tr}[\mathcal{E}(\widetilde{\rho}(g))|\psi(g)\rangle\langle\psi(g)|\otimes
\mathds{1}]$, is what we mainly considered. Here, $\mathds{1}$
denotes the Identity in single-qubit Hilbert Space. In the
following, we will search for all kinds of POVM measurements and
derive the optimal Information-Disturbance tradeoff.

To give a precise meaning to our problem, we need to investigate some properties of quantum measurement. According to Kraus's theory\cite{Kraus}, each
map $\mathcal{E}_r$ can be written in the form of operator decomposition $\mathcal{E}_r(\widetilde{\rho}(g))=\sum_\mu A_{r\mu}\widetilde{\rho}(g) A_{r\mu}^\dag$ and
will provide the state $\rho_r=\sum_\mu A_{r\mu}\widetilde{\rho}(g) A_{r\mu}^\dag/p(r|g)$ (normalized) after the result $r$ is observed, where, $p(r|g)=\sum_\mu  A_{r\mu}^\dag A_{r\mu}\widetilde{\rho}(g)$
 is the conditional probability of outcome $r$ occurring given that state $\widetilde{\rho}(g)$ is being input. The trace-Preserving condition for maps $\mathcal{E}_r$ further
 requires $\sum_{r\mu} A_{r\mu}^\dag A_{r\mu}=\mathds{1}\otimes\mathds{1}$.
For a completely unknown direction $\vec{n}$, one can assume the corresponding group parameter $g$ is chosen randomly with the uniform probability distribution. Thus, by averaging
over all the possible measuring outcome $r$ and pure state $|\psi(g)\rangle$, one can identify the average information and average disturbance
for the POVM $\{A_{r\mu}\}$ as follows:
\begin{eqnarray}
\mathcal{I}&=&\int_\mathbb{G} dg \sum_{r}p(r|g)|\langle
\psi_r|\psi_g\rangle|^2\nonumber\\
&=&\int_\mathbb{G} dg \sum_{r\mu}\mathrm{Tr}[
A_{r\mu}^{\dagger}A_{r\mu} \widetilde{\rho}(g)]|\langle
\psi_r|\psi_g\rangle|^2,\label{g1}\\
\mathcal{D}&=&1-\mathcal{F}\nonumber\\
&=&1- \int_\mathbb{G} dg \sum_{r\mu}\mathrm{Tr}[A_{r\mu}
\widetilde{\rho}(g) A_{r\mu}^{\dagger} |\psi(g)\rangle\langle
\psi(g)|\otimes \mathds{1}].\label{f1}
\end{eqnarray}

To be specific, we shall reuse the notation in Eq. (\ref{vecn}) and assign a rotation $r\in \mathrm{SU}(2)$ for the guessed state $|\psi_r\rangle$. By rewriting
$|\psi_r\rangle=U_r|0\rangle$, we can reduce the expression $\mathcal{I}$ to
\begin{eqnarray}
\mathcal{I}&=&\int_\Omega dg \sum_{r\mu}\mathrm{Tr}[
A_{r\mu}^{\dagger}A_{r\mu} \widetilde{\rho}(g)]|\langle
0|U_r^\dagger U_g|0\rangle|^2
\nonumber\\
&=&\int_\Omega dg^\prime \sum_{r\mu}\mathrm{Tr}\left[(U_r^{\dagger}\otimes U_r^{\dagger})A_{r\mu}^\dagger A_{r\mu}(U_r\otimes U_r)
\widetilde{\rho}(g^\prime)\right]\nonumber\\
&\times & \langle \psi_{g^\prime}|0\rangle\langle 0|\psi_{g^\prime}\rangle,\label{g2}
\end{eqnarray}
where in the second line we have defined
$U_{g^\prime}=U_r^{\dagger}U_g$ and apply the invariance $dg^{\prime}=dg$.

Generally, the value of information $\mathcal{I}$ varies greatly depending on the intensity of Eve's eavesdropping attack. However, two extreme cases have already been known:
(1) The most informative measurement, with $\mathcal{I}_{max}=\frac{3+\sqrt{3}}{6}$, happens when Von Neumann Projection along the four tetrahedral
directions is performed\cite{Gisin}. (2) $\mathcal{I}_{min}=2/3$ regards to the case inwhich the projective measurement is performed within the Hilbert Space of
$|\psi(g)^\perp\rangle$ only, leaving the first state $|\psi(g)\rangle$ intact, i.e., $\mathcal{D}(\mathcal{I}_{min})=0$ \cite{Mass95}. To escape from being detected, Eve may adjust
his strategy, varying his information from $\mathcal{I}_{max}$ to $\mathcal{I}_{min}$. In this case, what is Eve's minimal disturbance for each intermediate information  $\mathcal{I} ~( \mathcal{I}_{min}\le
\mathcal{I}\le \mathcal{I}_{max})$
is what we mainly focused and is in fact the mathematic description of the tradeoff problem.

%To derive the optimal trade-off between the information gain $I$ and the state disturbance $D$,
For this purpose, one needs to perform an exhaustive examination of
all the possible $\{A_{r\mu}\}$. Fortunately, one can resort to the
group covariant technique to strikingly simplify our problems.

\section{Covariant Measurement and Optimal Information-Disturbance Bound}\label{bound}

Group covariant quantum measurement is a special kind of measurement which originates from the symmetry of the input state and has already been
proven to be optimal in quantum state estimation \cite{qes} and the quantum cloning \cite{qclo} process. It can be easily shown that
the optimality also preserves in our problem. In fact, for an arbitrary (covariant or non-covariant)
CP map $\mathcal{E}(\rho)=\sum_{r\mu}A_{r\mu}\rho A_{r\mu}^{\dagger}$, one can
construct a covariant CP map $\mathcal{E}^\prime(\rho) =\int_h\mathcal{E}_h^\prime(\rho) dh$ which yields the same amount of information gain and disturbance, when the
$\mathcal{E}_h^\prime(\cdot)$ and the guessed state for result $h$ are chosen to be
\begin{eqnarray}
\mathcal{E}_h^\prime(\rho) &=&  \sum_{r\mu}(U_h U_r^{\dagger}\otimes U_h
U_r^{\dagger})A_{r\mu}(U_r U_h^{\dagger}\otimes U_r
U_h^{\dagger})\rho\nonumber \label{eh}\\
&&  \times (U_h U_r^{\dagger}\otimes U_h U_r^{\dagger})A_{r\mu}^\dag (U_r
U_h^{\dagger}\otimes U_r U_h^{\dagger}),\\
|\psi_h\rangle &=& U_h|0\rangle, \label{psih}
\end{eqnarray}
with the subscript $h\in SU(2)$ denoting the measurement result of the continuous POVM.
Therefore, the optimal trade-off bound for covariant map is also
the optimal bound for arbitrary maps. Therefore, in looking for the optimal bound between $\mathcal{I}$ and $\mathcal{}D$, there will be no
loss of generality if we restrict our study in the covariant way. The covariance map in Eq.(\ref{eh}) and (\ref{psih}), along with its good property
$\mathcal{E}_{gh}^\prime(\rho)=U_g\mathcal{E}_h(U_g^\dag\rho U_g)U_g^\dag$, not only guarantees the measurement achieves its optimal performance for all the
possible state $|\psi(g)\rangle$, but also simplifies our following computation considerably. Hereafter, we will consider the covariant instrument
\begin{eqnarray}
A_{h}=U_{h}\otimes U_{h}A_{0}U_{h}^{\dag }\otimes
U_{h}^{\dag} \label{Ah}
\end{eqnarray}
with the operator $A_0$ denoting a seed of the whole set of Kraus operators. Notice that the trace-preserving condition now boils down to
$\int_h A_{h}^\dag A_{h}=\mathds{1}\otimes \mathds{1}$ which can be further reduced with Schur's lemma for reducible group representation
\cite{zelo}:
\begin{eqnarray}
& & \int_{SU(2)} dgU_{h}\otimes U_{h}A_{0}^{\dagger}A_0 U_{h}^{\dag }\otimes U_{h}^{\dag }\nonumber \\
&=&\mathrm{Tr}[A_0^{\dagger}A_{0}\mathcal{M}_1]\mathcal{M}_1+\mathrm{Tr}[A_0^{\dagger}A_{0}\mathcal{M}_2]\mathcal{M}_2
/3,
\end{eqnarray}
where $\mathcal{M}_1=|\Psi^-\rangle\langle\Psi^-|$($|\Psi^\pm\rangle=(|01\rangle\pm|10\rangle)/\sqrt{2}$) denotes the
uni-dimensional completely asymmetric subspace and
$\mathcal{M}_2=\mathds{1}\otimes \mathds{1}-\mathcal{M}_1$ denotes the 3-dimensional
symmetric subspace. Now the trace-preserving condition boils down to
\begin{eqnarray}
\mathrm{Tr}[A_{0}^{\dagger}A_{0}\mathcal{M}_1]=1~~~ \text{and} ~~~~\mathrm{Tr}[A_{0}^{\dagger}A_{0}\mathcal{M}_2]=3.\label{trace}
\end{eqnarray}

With the covariant map $\{A_h\}$, the integral $dg$ in Eq.(\ref{g1})
and (\ref{f1}) can be easily obtained. For $\mathcal{D}$, we have
\begin{eqnarray}
\mathcal{D}&=&1- \int_\mathbb{G} dg\int_\mathbb{G} dh \mathrm{Tr}\left[A_h \widetilde{\rho}(g) A_h^\dag |\psi(g)\rangle\langle \psi(g)|\otimes I \right]\nonumber\\
&=&1-\sum_{i=0,1}\int_\mathbb{G} dg \langle \psi(g)|\langle i|A_0 \widetilde{\rho}(g)
A_0^{\dagger} |\psi(g)\rangle|i\rangle\nonumber\\
&=&1- \sum_{i,j,k=0,1}\int_\mathbb{G} dg \langle \psi(g)|\left(|j\rangle\langle j|\right)\langle i|A_{0} \widetilde{\rho}(g)
A_{0}^{\dagger}\nonumber\\
& &\times|k\rangle\langle k|\psi(g)\rangle|i\rangle \nonumber\\
&=&1-\sum_{i,j,k=0,1}\langle
ji|A_{0}M_{jk}A_{0}^{\dagger}|ki\rangle,\label{f2}
\end{eqnarray}where the operator
\begin{eqnarray}
M_{jk}=\int_\mathbb{G} dg \langle\psi(g)|j\rangle \cdot\widetilde{\rho}(g)\cdot \langle
k|\psi(g)\rangle, ~~ j,k\in\{0,1\}
\end{eqnarray}
can be calculated explicitly\cite{Mjk}.

The derivation for $\mathcal{I}$ can be done in a similar way, which yields
\begin{eqnarray}
\mathcal{I}&=&\int_\mathbb{G} dg \int_\mathbb{G} dh\mathrm{Tr}[
A_{h}^{\dagger}A_{h} \widetilde{\rho}(g)]|\langle
0|U_h^\dagger U_g|0\rangle|^2
\nonumber\\
&=&\int_\mathbb{G} dg\mathrm{Tr}\left[A_{0}^\dagger A_{0 }\widetilde{\rho}(g^\prime)\right]\cdot\langle \psi(g)|0\rangle\langle 0|\psi(g)\rangle\nonumber\\
&=&\mathrm{Tr}\left[A_{0}^\dagger A_{0}M_{00}\right].\label{g2}
\end{eqnarray}

Now putting all these results together, the tradeoff problem can be formulated with the following semi-definite programming problem:
\begin{eqnarray}
Min:\mathcal{D}(\mathcal{I})=1-\mathcal{F} \label{optimization}
\end{eqnarray}%
such that
\begin{eqnarray}
& & \mathcal{F}=\sum_{i,j,k=0,1}\langle
ji|A_{0}M_{jk}A_{0}^{\dagger}|ki\rangle,\nonumber\\
& & \mathcal{I}=\mathrm{Tr}\left[A_{0}^\dagger A_{0}M_{00}\right], A_0^\dag A_0\ge 0,\nonumber \\
& & \mathrm{Tr}[A_{0}^{\dagger}A_{0}\mathcal{M}_1]=1~~ \text{and} ~~\mathrm{Tr}[A_{0}^{\dagger}A_{0}\mathcal{M}_2]=3.
\end{eqnarray}%
Due to the complication of the minimization above\cite{FG}, an
analytical solution to is not always obvious and available. However,
in the rest, one will see that we can rely on the vector analysis
technique and derive the optimal tradeoff bound.
%Then, by showing that an exemplary measurement does
%exist and achieve such a bound, we obtain the optimal tradeoff.

To continue our discussion, we need to introduce a few
vectors $\{\vec{\mathbf{v}}_i=\{v_{i1},v_{i2}\}^\mathrm{T}, v_{ij}\in\mathbb{C}\}$, $(i=1,2,\cdots, 8,j=1,2)$ such that
\begin{eqnarray}
A_0=\left(
\begin{array}{c|c|c|c}
a_{11} & a_{12} & a_{13} & a_{14} \\
a_{21} & a_{22} & a_{23} & a_{24} \\ \hline
a_{31} & a_{32} & a_{33} & a_{34} \\
a_{41} & a_{42} & a_{43} & a_{44}%
\end{array}
\right)=\left(
\begin{array}{c|c|c|c}
\vec{\mathbf{v}}_1 & \vec{\mathbf{v}}_2 & \vec{\mathbf{v}}_3 &\vec{\mathbf{v}}_4\\
\hline \vec{\mathbf{v}}_5 & \vec{\mathbf{v}}_6 & \vec{\mathbf{v}}_7
&\vec{\mathbf{v}}_8
\end{array}
 \right).
\end{eqnarray}
This helps to give much simpler expressions to our problem. First of all, the trace-preserving equation (\ref{trace}) can be reduced to
\begin{eqnarray}
\sum_i |\vec{\mathbf{v}}_i|^2=4, \label{vi}\\
 |\vec{\mathbf{v}}_2-\vec{\mathbf{v}}_3|^2+
|\vec{\mathbf{v}}_7-\vec{\mathbf{v}}_6|^2=2.\label{v2376}
\end{eqnarray}
Then, it can be easily obtained that
\begin{eqnarray}
\mathcal{F}=\frac{1}{2}+\frac{1}{12}f, ~~~~~\mathcal{I}=\frac{1}{2}+\frac{1}{12}g,
\end{eqnarray}
with $f$ and $g$ defined by
\begin{eqnarray}
f&=&|\vec{\mathbf{v}}_2|^2-|\vec{\mathbf{v}}_3|^2+|\vec{\mathbf{v}}_7|^2-|\vec{\mathbf{v}}_6|^2-|\vec{\mathbf{v}}_1|^2-|\vec{\mathbf{v}}_8|^2\nonumber\\
&+&|\vec{\mathbf{v}}_7-\vec{\mathbf{v}}_6+\vec{\mathbf{v}}_1|^2+|\vec{\mathbf{v}}_8+\vec{\mathbf{v}}_2-\vec{\mathbf{v}}_3|^2-2,\label{frelation}\\
g&=&|\vec{\mathbf{v}}_2|^2-|\vec{\mathbf{v}}_3|^2+|\vec{\mathbf{v}}_6|^2-|\vec{\mathbf{v}}_7|^2\label{grelation}.
\end{eqnarray}

The optimization in Eq. (\ref{optimization}) can now be equivalently reduced to looking for a set of vectors $\vec{\mathbf{v}}_i$
that satisfy the constraints Eq.(\ref{vi}) (\ref{v2376}) and maximize $f$ for a given value $g$.

After some lengthy but not very interesting algebra, one can checked that the relation between $f$ and $g$ actually follows
\begin{eqnarray}
f\le f_{max}(g)=g+\sqrt{24-2g^2}. \label{fg}
\end{eqnarray}
This means that for any quantum measurement, the amount of the disturbance $\mathcal{D}$ caused on the quantum states must follows
\begin{equation}
\mathcal{D}\geq \mathcal{D}_{min}=1-\mathcal{I}-\sqrt{-\frac{1}{3}+2\mathcal{I}-2\mathcal{I}^2}.  \label{DItradeoff}
\end{equation}

In the literature, the quantum measurement whose disturbance equals
$\mathcal{D}_{min}$ is named as \textquotedblleft Minimal
Disturbance Measurement (MDM) \textquotedblright. This is the best
strategy for Eve, as it maximizes his information gain for a given
average disturbance.

The MDM quantum operation for antiparallel spins can be deduced from the derivation of Eq.(\ref{fg}). Here we omit the complicated process and list the main result.
In fact, the operators $A_0$ with
\begin{eqnarray}
\frac{\vec{\mathbf{v}}_2}{|\vec{\mathbf{v}}_2|}=\frac{\vec{\mathbf{v}}_3}{|\vec{\mathbf{v}}_3|}=\frac{\vec{\mathbf{v}}_8}{|\vec{\mathbf{v}}_8|},
\vec{\mathbf{v}}_1=\vec{\mathbf{v}}_4=\vec{\mathbf{v}}_5=\vec{\mathbf{v}}_6=\vec{\mathbf{v}}_7=\vec{\mathbf{0}}
\end{eqnarray}
is one example at hand. Particularly, to see the interpolation
between the two extreme cases mentioned in Sec. II, we can introduce
a control parameter $\theta$:
\begin{eqnarray}
& & A_0=|00\rangle\langle\Psi^-|+\frac{\sqrt{6}\cos\theta}{2}|00\rangle\langle
\Psi^+|+\sqrt{3}\sin\theta|10\rangle\langle 11|,\nonumber\\
& &\text{with}~~~~~\theta\in\left[0,\arccos (1/\sqrt{3})\right].
\label{Aoptimal}
\end{eqnarray}
It is straightforward to verify that, from equations (\ref{f2}) (\ref{g2}), that the performance of the covariant measurement Eq.(\ref{Aoptimal}) follows
\begin{eqnarray}
\mathcal{I}&=&\frac{1}{2}+\frac{\sqrt{3}}{6}\cos\theta, \\
\mathcal{D}&=&\frac{1}{2}-\frac{\sqrt{3}\cos\theta}{6}-\frac{\sqrt{6}\sin\theta}{6},
\end{eqnarray}
and the equality sign in Eq. (\ref{DItradeoff}) actually can be satisfied.

\begin{figure}
  \includegraphics[width=8.0cm]{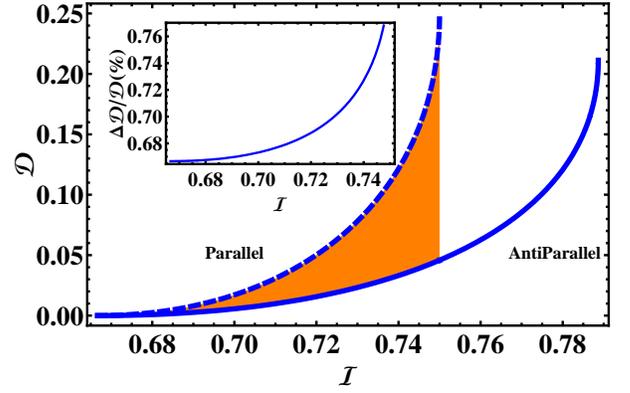}\\
  \caption{Comparative plot of information disturbance tradeoff between AntiParallel(Solid line, Eq.(\ref{DItradeoff})) and Parallel spins(Dashing line, from Ref.\cite{Nin1out}).
  Inset plot: the monotonically increasing reduction in disturbance $\Delta \mathcal{D}$, with the information gain varying from $0$ to $3/4$.
  The reduction oscillates between $66\%$ and $85\%$, and  maximum $\Delta \mathcal{D}/\mathcal{D}=\sqrt{2/3}$ is reached at $\mathcal{I}=3/4$.}\label{figg}
\end{figure}

\section{Discussions and concluding remarks} \label{Conclu}

Before concluding this work, we have two problems to remark.

The first one is the implementation of the covariant using only discrete POVMs. From the covariant operators above, we can construct
an POVM with only four outcomes. Other operator can be obtained by $A_i=\frac{1}{2}U_i\otimes U_i A_0
U_i^{\dagger}\otimes U_i^{\dagger}, ~~ (i=0,1,\dots,3)$, with $U_i$:
\begin{eqnarray}
U_0=I,U_1=\frac{\sqrt{3}}{3}\mathds{1}-i\sigma_y, \\
U_2=\frac{\sqrt{3}}{3}\mathds{1}+i\frac{\sqrt{6}}{6}\sigma_y+i\frac{\sqrt{2}}{2}\sigma_x,\\
U_3=\frac{\sqrt{3}}{3}\mathds{1}    +i\frac{\sqrt{6}}{6}\sigma_y-i\frac{\sqrt{2}}{2}\sigma_x.
\end{eqnarray}
One can easily checked that all these operator satisfies the
normalization condition $\sum_i A_i^{\dagger}A_i=\mathds{1}\otimes
\mathds{1}$ and the optimal tradeoff for measuring
antiparallel states follows Eq.(\ref{DItradeoff}). This indicates
the relation in Eq.(\ref{fg}) is exactly a tight one and cannot be
further improved any more.

The second one is the physical meaning of the tradeoff bound in Eq.
(\ref{DItradeoff}). We remark that it sheds some new lights on the
secure transmission of reframe information. Although the measurement
leading to maximal information have already been proposed
\cite{Gisin} and even experimentally implemented\cite{expFrame}, the
state disturbance after their measurement is $\mathcal{D}=1/4$,
which is not a minimal disturbance measurement for the antiparallel
spins, as shown in this paper. This means that there exists a much
better eavesdropping strategy for Eve. For example, by choosing
$A_0=|00\rangle (\sqrt{2+\sqrt{3}}\langle
01|+\sqrt{2-\sqrt{3}}\langle 10|)$, it is not difficult to show that
the minimum disturbance $\mathcal{D}=(3-\sqrt{3})/6$ can be reached.
This is far from a piece of good news for transmission information
via the antiparallel spins. With the antiparallel state, Alice gains
a definitely improvement in the precision or fidelity in her
transmission of frame information to Bob. But the precision or
fidelity is improved for Eve, too. what is worse, compared with the
parallel quantum state, Eve could obtain the same amount of
information with a less disturbance. To see this, in Fig.
\ref{figg}, we give a comparative plot of the
information-disturbance tradeoff between the case of antiparallel
spin and of the parallel spin. The tradeoff bound for Parallel spins
marked with Dashing line is borrowed from Ref.\cite{Nin1out}. It can
be obviously observed that antiparallel spins provides an
unexceptional improvement in the information gain ($\mathcal{I}$ up
to $(3+\sqrt{3})/6$). However, for the values $\mathcal{I}\leq 3/4 $
a pronounced decrease in the disturbance will be spotted. In order
to see the degree of decrease in a better way, we also plots the
dependence of decrease
$\Delta\mathcal{D}=\mathcal{D}_{anti}-\mathcal{D}_{para}$ on the
information gain. Numerical analysis reveals that the amount of
reduction in $\mathcal{D}$ increases monotonically with the gain,
with the maximum $\Delta\mathcal{D}/\mathcal{D}=\sqrt{2/3}=81.65\%$
attained at $\mathcal{I}=3/4$.

In conclusion, we give a heuristic security analysis of transmitting
reference reframes, with the model of information-disturbance
tradeoff. A strict bound for antiparallel spins, along with the
optimal POVM measurement which attains the bound is obtained.
Finally, we give a comparison between the tradeoff in antiparallel
and parallel cases, which reveals that the improvement in
information gain doesn't always mean a good matter, at least in the
cases when information is being measured with fidelities. We believe
more thorough analysis using the information theory-based methods
should be required.

\begin{acknowledgments}
This work was supported by National Fundamental Research Program, also by
National Natural Science Foundation of China (Grant No. 10674128 and
60121503) and the Innovation Funds and \textquotedblleft Hundreds of
Talents\textquotedblright\ program of Chinese Academy of Sciences and Doctor
Foundation of Education Ministry of China (Grant No. 20060358043)
\end{acknowledgments}


\begin{thebibliography}{99}

\bibitem{RMP} S. D. Bartlett, T. Rudolph, R. W. Spekkens, Rev. Mod.
Phys. \textbf{79}, 555 (2007).
\bibitem{pereUnspeak} A. Peres and Petra F. Scudo, Arxiv: quant-ph/0201017.
\bibitem{pere91} A. Peres and W. K. Wootters, Phys. Rev. Letts
\textbf{66}, 1119 (1991).

\bibitem{Nspin1} E. Bagan, M. Baig, and R. Mu\~{n}oz-Tapia, Phys.
Rev. Lett. \textbf{87}, 257903 (2001);\textit{ibid}, \textbf{89},
277904 (2002); Phys. Rev. A \textbf{70}, 030301 (2004).

\bibitem{Nspin2} E. Bagan, M. Baig, A. Brey, and R. Mu\~{n}oz-Tapia,
R. Tarrach, Phys. Rev. Lett. \textbf{85}, 5230 (2000).

\bibitem{Nspin3} G. Chiribella, G. M. D'Ariano, P. Perinotti, and M. F. Sacchi,
Phys. Rev. Lett. \textbf{93}, 180503 (2004).
\bibitem{pere01carte} A. Peres and Petra F. Scudo,
% Transmission of a Cartesian Frame by a Quantum System,
 Phys. Rev. Lett. \textbf{87}, 167901 (2001).

\bibitem{pere01} A. Peres and Petra F. Scudo,
% Entangled Quantum States as Direction Indicators,
Phys. Rev. Lett. \textbf{86} 4160(2001).

\bibitem{Mass95} S. Massar and S. Popescu,% Optimal extraction of
%Information from Finite Quantum Ensembles,
Phys. Rev. Lett. \textbf{95}, 1259 (1995).
\bibitem{Gisin} N. Gisin, S. Popescu, Phys. Rev. Lett. \textbf{83}, 432 (1999).

\bibitem{Nspin4} Piotr Kolenderski and Rafal Demkowicz-Dobrzanski, Phys. Rev. A \textbf{78}, 052333 (2008).
\bibitem{Chir} Giulio Chiribella, Lorenzo Maccone, and Paolo
Perinotti, Phys. Rev. Lett. \textbf{98}, 120501 (2007).


\bibitem{fuchs96.pra} C. A. Fuchs and A. Peres, Phys.\ Rev. A \textbf{53},
2038 (1996).
\bibitem{fuchs01.pra} C. A. Fuchs and K. Jacobs, Phys. Rev. A \textbf{\ 63},
062305 (2001).
\bibitem{mista05.pra} L. Mi{\v{s}}ta Jr., J. Fiur\'{a}{\v{s}}ek, and R.
Filip, Phys. Rev A \textbf{72}, 012311 (2005).
\bibitem{macca} L. Maccone, Phys. Rev. A \textbf{73}, 042307 (2006).

\bibitem{discrimiVS} Francesco Buscemi and Massimiliano F. Sacchi , Phys.\
Rev. A \textbf{74} 052320 (2006).

\bibitem{banaszek01.prl} K.~Banaszek, Phys. Rev. Lett \textbf{86}, 1366 (2001).
\bibitem{banaszek00.pra} K.~Banaszek, Phys. Rev. A \textbf{62}, 024301 (2001).

\bibitem{M.F.Sacchi} M. F. Sacchi, Phys. Rev. Lett \textbf{96}, 220502 (2006).
\bibitem{Davies} E. B. Davies, IEEE Trans. Info. Theo. IT, \textbf{24}, 596
(1978).
\bibitem{Kraus} K. Kraus, \emph{States, Effects, and Operations},
(Springer-Verlag, Berlin, 1983).


%%%%%%%%%%%%%%%%%%%%%%%%%%%

\bibitem{qes} G. M. D'Ariano, and M. F. Sacchi, Phys. Rev. A \textbf{72},
042338 (2005).

\bibitem{qclo} G. Chiribella, G. M. D'Ariano, and P. Perinotti Phys. Rev. A
\textbf{72}, 042336 (2005).

\bibitem{zelo} D. P. Zhelobenko, \emph{Compact Lie Groups and Their
Representations} (American Mathematical Society, Providence, RI, 1973).

%\bibitem{expFrame} Evan R. Jeffrey, Joseph B. Altepeter, Madalina Colci, and Paul G.
%Kwiat,
% "Optical Implementation of Quantum Orienteering",
%Phys.Rev. Lett. \textbf{96} 150503 (2006)



\bibitem{Mjk} With a similar technique in Ref.\cite{banaszek00.pra}, it is not difficult to calculate the matrix entries for $M_{jk}$. Define $|\Psi^{\pm}\rangle=(|01\rangle\pm|10\rangle)/\sqrt{2}$, the operators $M_{jk}$ can be represented by $
M_{00}=\left(|0\rangle\langle 0|\otimes \mathds{1}+\mathds{1}\otimes
|1\rangle\langle 1|\right)/12+|\Psi^-\rangle\langle\Psi^-|/6,
M_{11}=\left(\mathds{1}\otimes |0\rangle\langle 0|+|1\rangle\langle
1|\otimes \mathds{1}\right)/12+|\Psi^-\rangle\langle\Psi^-|/6,
M_{01}=M_{10}^\dagger=\left(|\Psi^-\rangle\langle
11|-|00\rangle\langle\Psi^-|\right)/6\sqrt{2}. $
\bibitem{FG} Although the infomration gain can be finally written as
\begin{eqnarray}
\mathcal{I}&=&\frac{1}{3}+\frac{1}{6\sqrt{2}}\left(\langle
\Psi^-|A_0^{\dagger}A_0|01\rangle+\langle
01|A_0^{\dagger}A_0|\Psi^-\rangle\right), \nonumber
\end{eqnarray}
there seems to be no simpler formula for $\mathcal{F}$
\begin{eqnarray}
\mathcal{F}&=&I+\frac{1}{6}\left[\sum_i|\langle 1i|A_0|10\rangle
|^2-\sum_i|\langle 1i|A_0|01\rangle |^2\right]\nonumber\\
&+&\frac{1}{6\sqrt{2}}(\mathrm{Tr} [A_0|\psi^-\rangle\langle
11|A_0^{\dagger}|1\rangle\langle 0|\otimes \mathds{1}]-\nonumber\\
 & & \mathrm{Tr} [A_0|00\rangle\langle \psi^-|A_0^{\dagger}|1\rangle\langle 0|\otimes \mathds{1}]+c.c.),\nonumber
\end{eqnarray}
inwhich $c.c.$ denotes \emph{Complex Conjugate}.

\bibitem{expFrame} Evan R. Jeffrey, Joseph B. Altepeter, Madalina Colci, and Paul G.
Kwiat, % "Optical Implementation of Quantum Orienteering",
Phys.Rev. Lett. \textbf{96} 150503 (2006)


\bibitem{Nin1out} Ladislav Mi\v{s}ta. Jr., Jarom\'{\i}r Fiur\'{a}\v{s}ek.
Phys.\ Rev. A \textbf{74}, 022316 (2006).

%\bibitem{gmd} G. M. D'Ariano, Fortschr. Phys \textbf{51}, 318 (2003).

%\bibitem{bb84} C. H. Bennett and G. Brassard, in \emph{Proceedings of the
%IEEE International Conference on Computers, Systems, and Signal
%Processing, Bangalore, India} (IEEE, New York, 1984), pp. 175;

%\bibitem{E91} A. K. Ekert, Phys. Rev. Lett. \textbf{67}, 661 (1991).



\end{thebibliography}
\end{document}